# Attosecond Timing in Optical-to-Electrical Conversion

Fred N. Baynes[1], Franklyn Quinlan[1], Tara M. Fortier[1], Qiugui Zhou[2], Andreas Beling[2], Joe C. Campbell[2], Scott A. Diddams[1]

1Time and Frequency Division, National Institute of Standards and Technology, Boulder, CO 80305 USA
2Department of Computer and Electrical Engineering, University of Virginia, Charlottesville, Virginia 22904 USA
*Corresponding author: frederick.baynes@nist.gov, franklyn.quinlan@nist.gov

**The most frequency-stable sources of electromagnetic radiation are produced optically, and optical frequency combs provide the means for high fidelity frequency transfer across hundreds of terahertz and into the microwave domain. A critical step in this photonic-based synthesis of microwave signals is the optical-to-electrical conversion process. Here we show that attosecond (as) timing stability can be preserved across the opto-electronic interface of a photodiode, despite an intrinsic temporal response that is more than six orders of magnitude slower. The excess timing noise in the photodetection of a periodic train of ultrashort optical pulses behaves as flicker noise (1/f) with amplitude of 4 as/√Hz at 1 Hz offset. The corresponding fractional frequency fluctuations are $1.4 \times 10^{-17}$ at 1 second and $5.5 \times 10^{-20}$ at 1000 seconds. These results demonstrate that direct photodetection, as part of frequency-comb-based microwave synthesis, can support the timing performance of the best optical frequency standards, and thereby opens the possibility for generating microwave signals with significantly better stability than any existing source.**

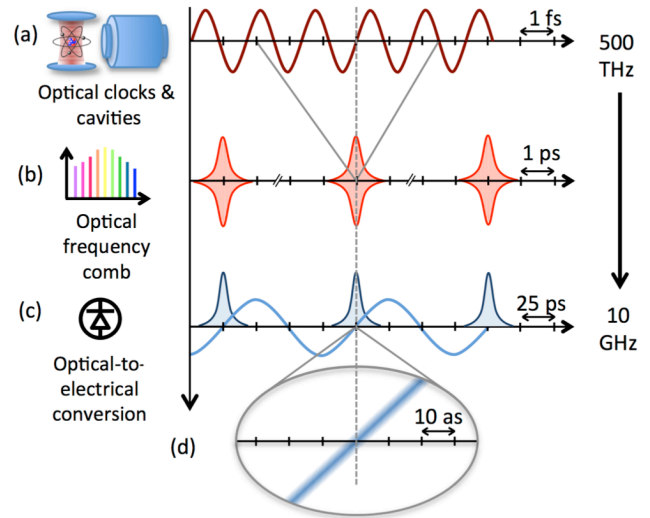

**Fig. 1.** Conceptual diagram of optical frequency division illustrating the different time scales for the optical and microwave signals. (a) An ultra-stable CW optical frequency reference, such as an optical cavity or optical atomic clock, provides the intrinsic timing stability at and below the intrinsic femtosecond scale of the optical carrier wave [1-4]. (b) The optical timing stability is transferred to an optical frequency comb that provides a coherent link to the microwave-rate optical pulse train that consists of sub-picosecond pulses [10,11]. (c) Photodetection of the optical pulse train results in electrical pulses on the 10 ps timescale from which a 10 GHz microwave signal is obtained. In this work we explore the noise limitations of this optical-to-electrical conversion process and show that timing can be preserved down to the attosecond (as) level. As indicated in the expanded view of (d), this implies a time-averaged splitting of the 100 ps period of the photonically-generated microwaves by more than a factor of $10^7$.

## Introduction

Modern optical frequency standards have extraordinary performance, with lasers locked to passive reference cavities demonstrating fractional frequency instability as low as $\delta f/f = 1 \times 10^{-16}$ at 1 s [1], and stabilized lasers, either proposed or presently under development, could realize instabilities at or below $10^{-17}$ on the same 1 s timescale [1, 2]. Moreover, optical atomic clocks already approach instabilities of $\delta f/f = 1 \times 10^{-18}$ at $10^4$ s [3, 4]. This extremely fine subdivision of the period of an optical cycle could impact a broader range of applications if maintained in the microwave domain. Examples include precision microwave spectroscopy [5], synchronization at kilometer-scale facilities such as phased-array telescopes and x-ray free-electron lasers [6, 7], and remote sensing and radar systems [8]. Optical frequency division (OFD) with a mode-locked laser comb [9] provides a means to realize this goal, and its main features are shown in Fig. 1.

Many aspects of OFD have been studied in detail, and the residual stability of a frequency comb locked to an optical reference has been shown as less than $3 \times 10^{-17}$ at 1 second [10, 11]. Characterization of saturation and excess noise in high-speed photodetection [12-17] and the generation of 10 GHz microwave signals with jitter at the femtosecond level [9, 12] have also been undertaken. However, with the advent of even lower noise optical references, the ultimate precision with which an ultra-stable optical pulse train can be converted into an electronic signal becomes an important question. One approach to optical-to-electrical conversion, known as balanced optical-microwave photodetection (BOM-PD), uses a voltage controlled microwave oscillator that is phase locked to the stable optical pulse train through an optical fiber Sagnac interferometer [18-21] However, the simplicity of direct pulse detection remains attractive, and also makes available broad tunability and the use of any harmonic of the repetition rate within the detector bandwidth. Here we address the attosecond timing limits of direct photodetection by comparing 10 GHz microwave signals generated from two high-speed photodiodes illuminated by the same optical pulse train. In doing so, we demonstrate the lowest noise in optical-to-electrical noise achieved to date, measuring over 7 decades of offset frequencies, from 1 mHz to 10 kHz. The phase noise at 1 Hz offset from the 10 GHz carrier is near -135 dBc/Hz, falling to -170 dBc/Hz at 10 kHz. These values correspond to equivalent timing jitter amplitudes of 4 as/√Hz and 0.07 as/√Hz, respectively. Moreover, the fractional frequency instability was determined to be $1.4\times10^{-17}$ at one second and $5.5\times10^{-20}$ at 1000 seconds – consistent with the phase noise measurements. This level of performance is fully capable of supporting the best optical frequency references and generating electronic signals with higher stability than the best present microwave sources on this same time scale [22] appears achievable.

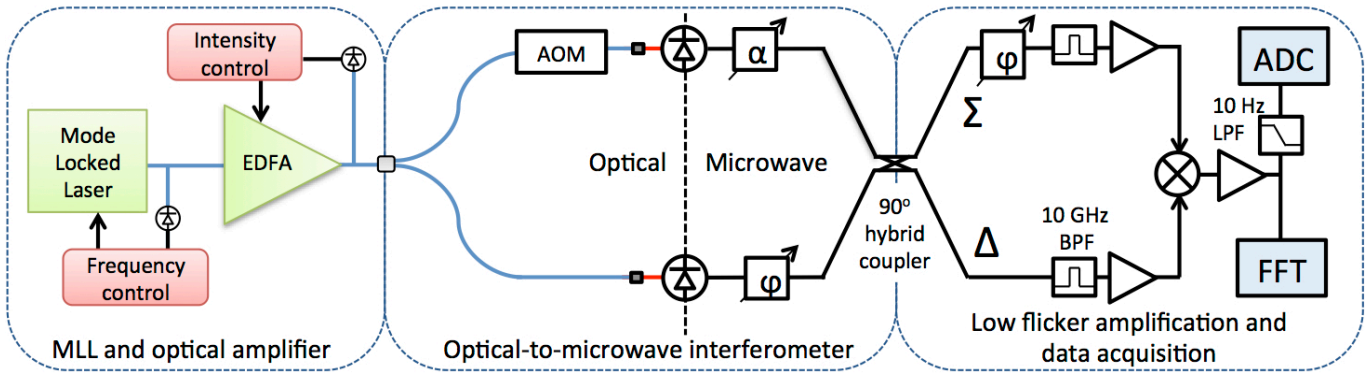

**Fig. 2.** Measuring optical-to-microwave conversion noise with the carrier suppression technique. MLL mode-locked laser, EDFA Erbium-doped fiber optical amplifier, AOM acousto-optic modulator, α variable attenuator, φ variable phase shifter, ADC analog to digital converter, FFT Fourier transform spectrum analyzer, BPF band pass filter, LPF low pass filter.

**Optical-to-Electrical Conversion Measurement**

To determine the timing fidelity of the optical-to-electrical conversion process, a phase comparison was performed on 10 GHz carriers generated by two short pulse-illuminated modified uni-traveling carrier (MUTC) photodetectors. The residual photodetection noise is comparable to the noise of the best microwave amplifiers and mixers, making the traditional saturated mixer measurement approach untenable. An interferometric carrier suppression technique was therefore used [17, 23, 24]. For this approach the optical pulse train from a mode-locked laser (MLL) was split to illuminate the two MUTC photodiodes, as shown in Fig. 2. The photodetected microwave signals were recombined electronically, forming an interferometer with one half in the optical domain and the other in the microwave domain. The phase and amplitude of the microwaves were set such that the two signals constructively interfere at the interferometer sum (Σ), or "bright" port, while destructive interference is obtained at the difference (Δ), or "dark" port. The output of the difference port consists of uncorrelated noise added by the two interferometer arms, dominated by the residual noise in the photodetection process. This noise was then amplified and mixed in quadrature with the output of the sum port. By suppressing the carrier in the difference port, close-to-carrier noise added by the amplifiers and mixer is greatly reduced [24], leading to a much lower noise floor of the measurement system for frequencies below 10 kHz. Throughout the measurement, a carrier suppression ratio greater than 60 dB was maintained, which required frequency and intensity control of the MLL. Intensity stabilization of the MLL ensured that slow intensity fluctuations, combined with the small difference in photodiode responsivity, did not disrupt the carrier suppression at the dark port. Voltage fluctuations at the output of the mixer were recorded with a Fast Fourier Transform (FFT) spectrum analyzer and a 10 Hz bandwidth analog-to-digital converter (ADC). After calibration, these measurements provide the power spectral density of phase fluctuations (from FFT) as well as a longer-term record (from ADC) of the phase and timing fluctuations. An acousto-optic modulator (AOM) in one arm of the interferometer was used to generate amplitude modulation for setting the mixer in the correct quadrature. The noise floor of the carrier suppression measurement system was ascertained by splitting a common signal from a commercial synthesizer into the microwave portion of the interferometer.

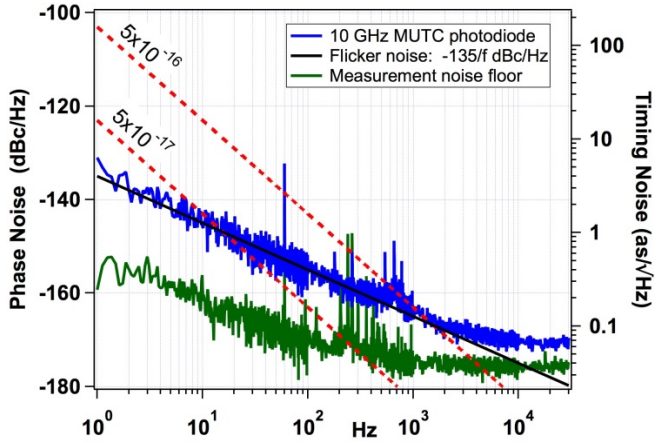

**Fig. 3.** The excess phase and timing noise spectrum for the generation of a 10 GHz microwave signal from direct photodetection. The solid black line is flicker noise of -135 dBc/Hz, and the lower green curve is the measurement noise floor. For comparison, the red dashed lines are the projected phase noise of ultra-stable CW lasers (divided to 10 GHz) having combined flicker and white frequency noise floors equivalent to fractional frequency stability of $5\times10^{-16}$ and $5\times10^{-17}$ at 1 s, respectively.

The two MUTC photodiodes used in this experiment are designed to have high microwave output power and small non-linearity to minimize amplitude-to-phase noise (AM-to-PM) conversion [25]. They are 40 μm in diameter and flip-chip bonded to enhance heat dissipation with one on a diamond substrate, the other on AlN, and both contacted to Peltier coolers for temperature stabilization and control. To improve photodiode linearity at 10 GHz, the 208.33 MHz repetition rate of the optical pulse train was increased to 3.33 GHz via an optical fiber pulse interleaver [15]. Both detectors were operated at 24 mA of photocurrent. Bias voltages of -18V and -12 V resulted in 10 GHz power of 12.6 dBm and 14 dBm, respectively. The bias voltages were chosen such that each photodiode was at a null of the AM-to-PM conversion [12, 13, 16].

### Results

The 10 GHz excess phase noise for a single MUTC detector is shown in Fig. 3, where 3 dB has been subtracted from the measurement under the assumption that both photodetectors contribute to the noise equally. The flicker noise is near -135 dBc/Hz at 1 Hz, and continues as 1/f over more than 3 decades to a white noise floor. When converted to timing noise density, the residual phase noise corresponds to 4 attoseconds/√Hz at 1 Hz offset and 70 zeptoseconds/√Hz at 10 kHz. The white phase noise floor is slightly above the measurement noise floor, and is attributed to uncorrelated optical amplifier noise between the two photodiodes that is not captured by the noise floor measurement.

A 6000 s measurement of the residual timing noise at 10 GHz was also recorded with the ADC and is shown in Fig. 4 along with the timing deviation calculated from these data [26]. At 1 second, the time deviation is only 4.2 attoseconds and remains below 12 attoseconds for timescales up to 100 seconds. The standard Allan deviation frequency instability was also calculated by differentiating the data of Fig. 4, and the result for a single MUTC detector is shown in Fig. 5. The fractional frequency instability at 1 second was found to be $1.4 \times 10^{-17}$ and decreases as nearly $1/\tau$. This is consistent with the theoretical calculation for flicker phase noise at -135 dBc/Hz [26], which is also plotted on Fig. 5. At longer averaging times of 1000 seconds the residual frequency instability ultimately reaches $5.5 \times 10^{-20}$.

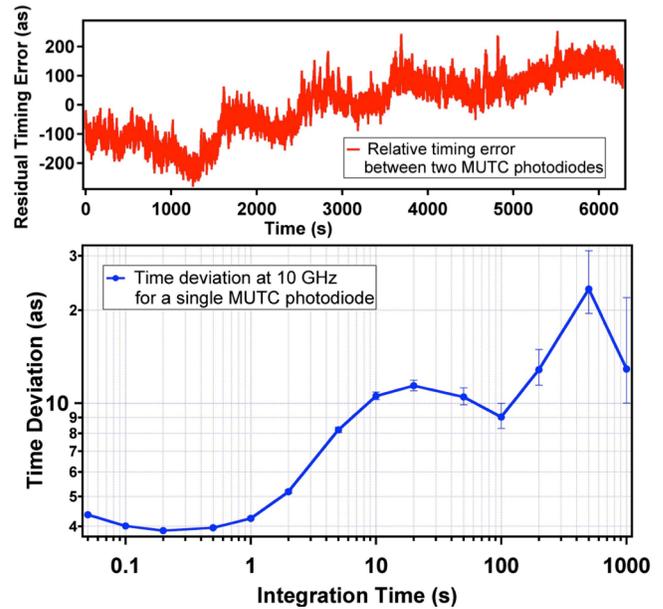

**Fig. 4.** (Top) The residual timing error in the optical-to-microwave conversion process for 2 MUTC photodiodes. The RMS variation over 6000 seconds of measurement is less than 106 attoseconds in a measurement bandwidth of 10 Hz. (Bottom) The timing deviation for a single MUTC photodiode for a 10 GHz carrier.

Since the carrier suppression measurement technique is nominally insensitive to phase noise arising from common-mode laser RIN, we have taken care to verify that any residual AM-to-PM would not increase the noise beyond our measurements. This is done by independently measuring the AM-to-PM conversion of the MUTC detectors, and the result is shown in Fig. 6. By appropriately adjusting the bias voltage, over 40 dB of AM-to-PM suppression could be obtained at a chosen photocurrent. Given the measured RIN from our Er:fiber laser comb [27], the RIN-induced phase noise is around -160 dBc/Hz at 1 Hz and -180 dBc/Hz at 10 kHz. Therefore, the added noise from AM-to-PM is negligible, and the data of Figs. 3-5 accurately represent the intrinsic phase, timing, and frequency noise in optical-to-electrical conversion with MUTC photodiodes.

### Discussion

Our measurements at 10 GHz have demonstrated the lowest phase, timing, and frequency noise in optical-to-electrical conversion of which we are aware. Over 1 second of measurement, corresponding to 10 billion cycles, the fractional frequency instability corresponds to only $10^{-7}$ of a cycle. The equivalent timing error is only $4\times10^{-18}$ s (4 attoseconds) nearly 6 orders of magnitude smaller than the 100 ps period of the 10 GHz signal.

It is instructive to compare the added noise in optical-to-electrical conversion to the relevant state-of-the-art optical and microwave frequency sources at different timescales. As shown in

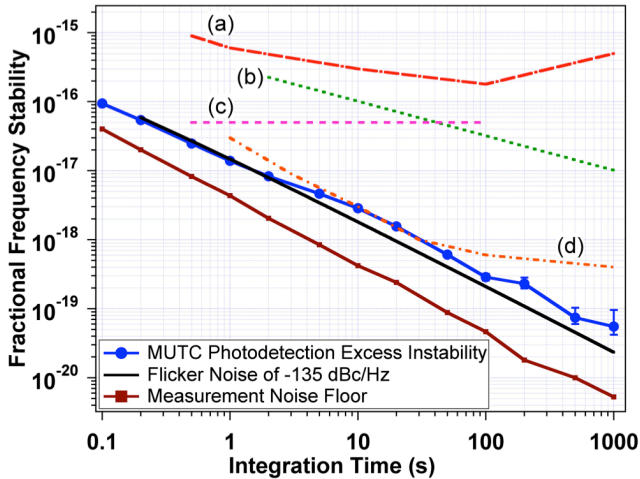 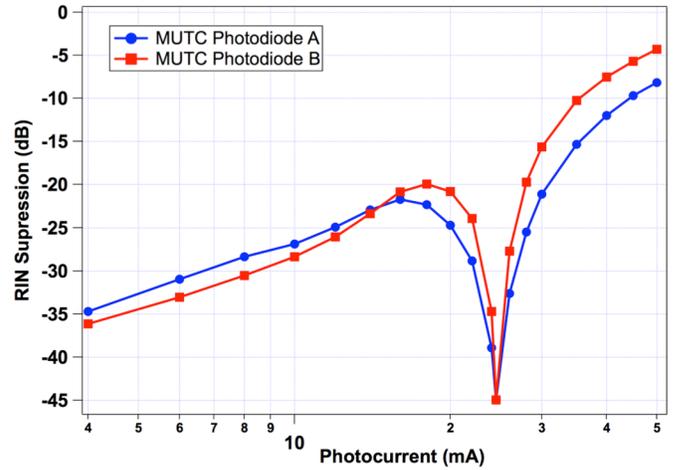

**Fig. 5.** The fractional frequency stability of the excess noise in direct photodetection leading to a 10 GHz microwave signal (10 Hz measurement bandwidth). The dashed lines are: (a) the stability of a cryogenic sapphire microwave oscillator [22], (b) neutral atom lattice clock [3][4], (c) the projected thermal noise floor of an optical silicon Fabry-Perot cavity [1], and (d) residual noise of an optical frequency comb [10]. Also shown are the projected flicker noise of -135 dBc/Hz and the measurement noise floor.

**Fig. 6.** The AM-to-PM coefficient of the two MUTC photodiodes used in this work expressed as the suppression of RIN converted to single sideband phase noise. The bias voltages -18 V (blue curve) and -12 V (red curve) were chosen for the two different photodiodes to provide an AM-to-PM null at 24 mA of photocurrent.

Fig. 5, the frequency stability results demonstrate that direct detection with MUTC detectors could preserve the fractional frequency stability of optical atomic clocks, as the instability added by photodetection is more than an order of magnitude below state-of-the-art Yb and Sr lattice clocks [3, 4]. Such long-term frequency stability requires continuously operating the MUTC photodiodes at an AM-to-PM null. As a measure of the AM-to-PM null stability, the photodiode temperature was changed 10 K, yielding no measurable change in the AM-to-PM coefficient. The temperature-to-phase coefficient of the MUTC photodiode was also measured by increasing the temperature by 10 K, and noting the change in the phase of the microwave signal. A result of $2\times10^{-4}$ rad/K (or equivalently 3 fs/K) was obtained, which is nearly a factor of 10 better than previous results [16], and approximately the same temperature sensitivity as 8 cm of standard optical fiber.

The highest stability microwave sources on the same timescales of Fig. 5 are currently cryogenic sapphire oscillators near 11 GHz, where fractional frequency instability below $10^{-15}$ has been demonstrated, reaching a minimum $< 2 \times 10^{-16}$ near 100 s [22]. The projection of the thermal-noise-limited stability of a laser locked to a silicon cavity is at (or below) $5\times10^{-17}$, which is also indicated in Fig. 5 [1]. Thus, the direct photodetection of the output of a frequency comb referenced to the combination of such a laser and an optical atomic clock could provide microwave signals with stability exceeding any existing microwave source on timescales of 1 to greater than 1000 s.

For shorter timescales, the most stable optical signals are provided by passive optical cavities. Figure 3 compares the phase noise of existing and next-generation optical cavities to the phase noise of optical-to-electrical conversion. For this comparison, perfect frequency division of the optical frequency to 10 GHz was assumed, and the fractional frequency instability at 1 second was taken to be $5\times10^{-16}$ for a typical state-of-the-art optical cavity and $5\times10^{-17}$ for a next-generation cavity. As this comparison shows, for the present state-of-the-art, the photodiode flicker noise would not have an impact until 1 kHz offset frequency. However, in the near future even flicker phase noise of only -135 dBc/Hz could limit the generation of 10 GHz microwaves at offset frequencies greater than 10 Hz. Extending the offset frequency range that can support next-generation optical cavities will require lowering the flicker noise of the photodetector. The origin of flicker noise in semiconductor devices is generally regarded to arise from conductivity fluctuations due to carrier traps, largely at surfaces and boundaries [28, 29][30]. How to lower flicker noise in semiconductor devices remains an important question in solid-state physics. However, for photodetectors with flicker noise dominated by surface effects, surface passivation has been shown to improve noise performance [29].

Optical-to-electrical conversion at the shortest timescales will be dominated by a white phase noise floor. In Fig. 3, the phase noise floor is believed to arise from optical amplifier noise, although ideally the phase noise would be shot noise limited. However, in the detection of ultrashort optical pulses, shot noise limited phase noise is usually well below the thermal noise floor [31]. Thus the circuit thermal noise, or in some cases photocarrier scattering [32], will yield the white phase noise limit in optical-to-microwave conversion. Despite being photodetector limited, it is worth noting that the white noise achievable with direct photodetection is comparable to state-of-the-art 10 GHz electronic sources [33] [34].

Optical frequency combs are a necessary part of the optical frequency division process, and will need the requisite low noise and high stability to take full advantage of the low noise capabilities of direct photodetection. As shown in Fig. 5, for timescales 1 second and longer, combs have demonstrated performance on par with the optical-to-electrical conversion stabilitly of direct photodetection [10, 11]. On timescales less than 1 second, the quantum-limited timing jitter of the frequency comb laser can be exceedingly low [35] such that more technical matters related to excess RIN in MLL pump lasers, servo bandwidth limitations, and the above-mentioned white noise floors at high offset frequencies tend to dominate [36].

## Summary

We report significant improvements in the excess phase, timing, and frequency noise in the optical-to-electrical conversion of an optical pulse train. For the generation of a 10 GHz carrier from direct photodetection of short optical pulses, the residual phase noise is nearly -135 dBc/Hz at 1 Hz and continues as flicker (1/f) to the white noise floor of -170 dBc/Hz at 10 kHz. The standard Allan deviation revealed a residual fractional frequency stability of $1.4\times10^{-17}$ at 1 second, and at longer timescales is consistent with flicker noise down to $5.5\times10^{-20}$ at 1000 seconds. The residual noise measured here indicates that, for timescales greater than $\sim 100$ ms, direct photodetection of a stable optical pulse train can support optical-to-microwave conversion of the highest stability optical sources, making optical stability available for a range of applications in the electronic domain. Shorter timescales are photodetection limited, but at a level comparable to the best electronic microwave sources. These results open the possibility for the photonic generation of microwave signals with unprecedented stability and low-noise.

## Funding Information

We thank Andrew Ludlow and Archita Hati for helpful comments on this manuscript, and Craig Nelson and Archita Hati for advice on the carrier suppression measurement. This work was supported by NIST and the DARPA PULSE program.